# Three-dimensional mid-infrared photonic circuits in chalcogenide glass

**A. Rodenas,[1,6] G. Martin,[2] B. Arezki,[2] N. D. Psaila,[3] G. Jose,[4] A. Jha,[4] L. Labadie,[5] P. Kern,[2] A. K. Kar,[1] and R. R. Thomson[1,7]**

[1] *Scottish Universities Physics Alliance (SUPA), School of Engineering and Physical Sciences, Heriot Watt University, Edinburgh EH14 4AS, United Kingdom*

[2] *UJF-Grenoble 1/CNRS-INSU, Institut de Planetologie et d'Astrophysique de Grenoble (IPAG), UMR 5274, Grenoble, France*

[3] *Optoscribe Ltd., 0/14 Alba Innovation Centre, Alba Campus, Livingston, Edinburgh EH54 7GA, United Kingdom*

[4] *Institute for Material Research, University of Leeds, Leeds LS2 9JT, United Kingdom*

[5] *I. Physikalisches Institut, Universität zu Köln, Zülpicher Str. 77, 50937 Köln, Germany*

[6] *e-mail: arodenas@gmail.com;* [7] *e-mail: r.r.thomson@hw.ac.uk*

We report the fabrication of single mode buried channel waveguides for the whole mid-infrared transparency range of chalcogenide sulphide glasses ($\lambda \leq 11$ µm), by means of direct laser writing. We have explored the potential of this technology by fabricating a prototype three-dimensional three-beam combiner for future application in stellar interferometry, which delivers a monochromatic interference visibility of 99.89% at 10.6 µm, and an ultrahigh bandwidth (3-11 µm) interference visibility of 21.3%. These results demonstrate that it is possible to harness the whole transparency range offered by chalcogenide glasses on a single on-chip instrument by means of direct laser writing, a finding that may be of key significance in future technologies such as astrophotonics and biochemical sensing.

The mid-infrared (MIR) range of the electromagnetic spectrum, between ~3 µm to ~30 µm wavelength, is a key region for a large number of applications in diverse areas such as biology and medicine [1], molecular spectroscopy [1,2], environmental monitoring [3], satellite remote sensing [4], and astronomy [5]. An essential step in order to unleash all of this potential science is to develop integrated optical platforms capable of addressing the technological requirements that each field demands. In this sense, the development of on-chip instruments such as optical sensors, high resolution spectrometers, or sophisticated beam combiners, is currently of high interest for the previous mentioned applications [1,2,5,6,7].

Although several MIR two-dimensional (2D) planar schemes have been recently developed [8,9], these are all based on multiple-step surface deposition and processing techniques, which place inherent limits to the device design and capabilities. In this Letter, we report the single-step fabrication of three-dimensional (3D) MIR photonic circuits inside chalcogenide glass, by means of ultrashort-pulse direct laser writing (DLW) [10-14]. We show that the MIR waveguide cores can be tailored in both size and refractive index, and can also be spatially positioned at will inside the material, making the chip extremely robust against mechanical stress, vibrations, humidity, and temperature changes [11]. Moreover, we also evidence that the useful range of these DLW waveguides is, as suspected [12], ultimately limited by the transparency range of the material used, and not by the fabrication technique.

In this work, high quality research and commercial chalcogenide sulphide glasses were used, both of which are free of highly toxic arsenic compounds. These glasses were commercial GaLaS (here after GLS) [14] and the research composition $75GeS_2$-$15Ga_2S_3$-4CsI-$2Sb_2S_3$-4SnS (here after GCIS) [15]. The MIR transmission upper limit of commercial GLS is known to be ~10 µm [14], while for GCIS we measured a slightly higher transmission upper limit of ~11 µm, as it is shown in Figure 1(a).

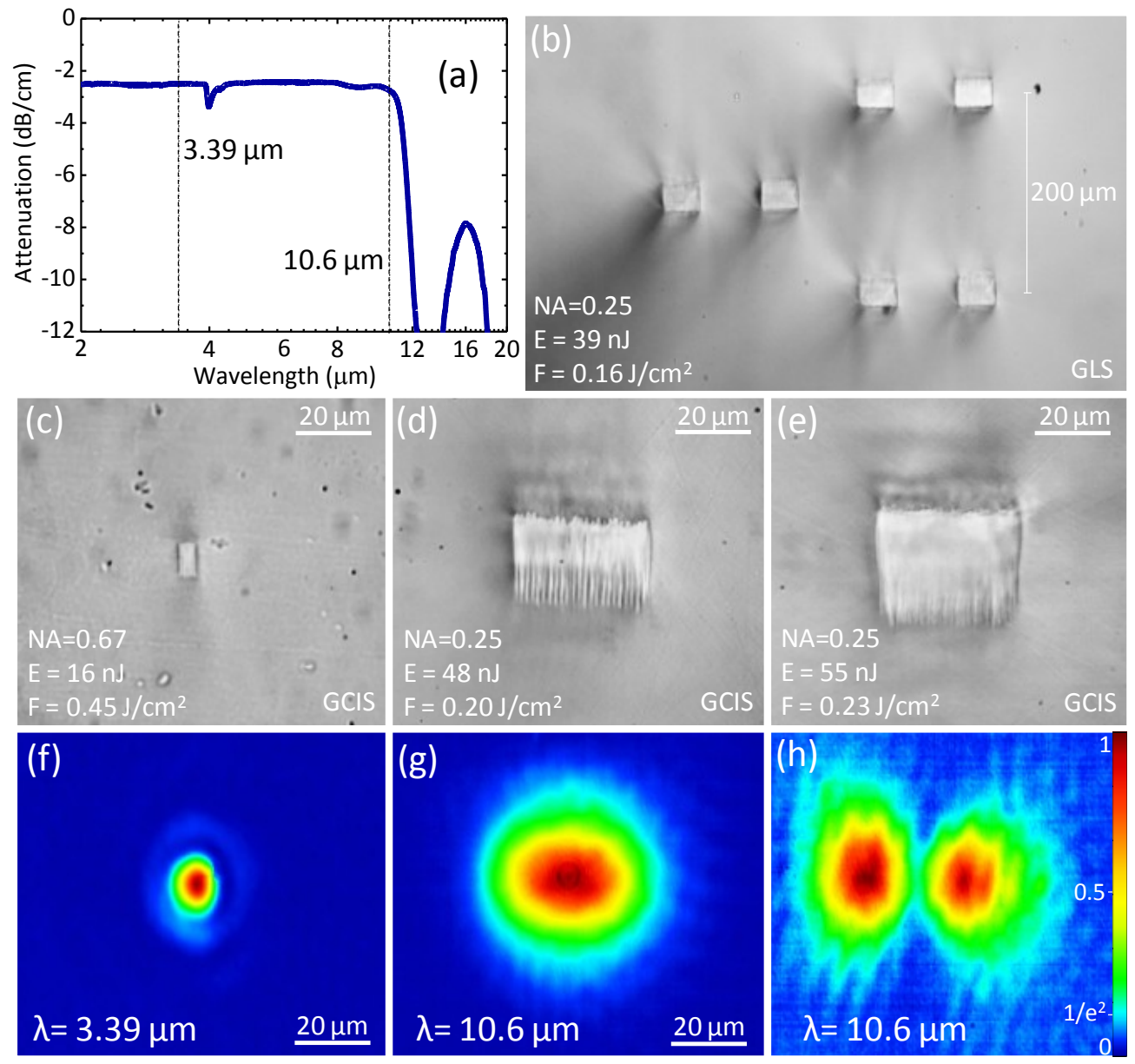

Fig. 1. (Color online) (a) Attenuation spectra of GCIS glass. (b) Example of a 2D waveguide array for the MIR. (c) Microscope image of a single mode waveguide at 3.39 µm. (d) and (e) Microscope images of single and bi-mode waveguides at 10.6 µm. (f), (g) and (h) Near-field mode images of waveguides shown in (c) (d) and (e) respectively.

Fig. 1(b) shows an example of a 2D waveguide array fabricated in GLS, showing the good waveguide core quality and 3D spatial positioning. Fig. 1(c) and (d) show cores fabricated in GCIS for single mode operation at 3.39 and 10.6 μm wavelengths, as seen in Figs. (f) and (g), respectively. Fig. 1(e) shows a bigger core which allowed the observation of higher order modes at 10.6 μm (Fig. 1(h)).

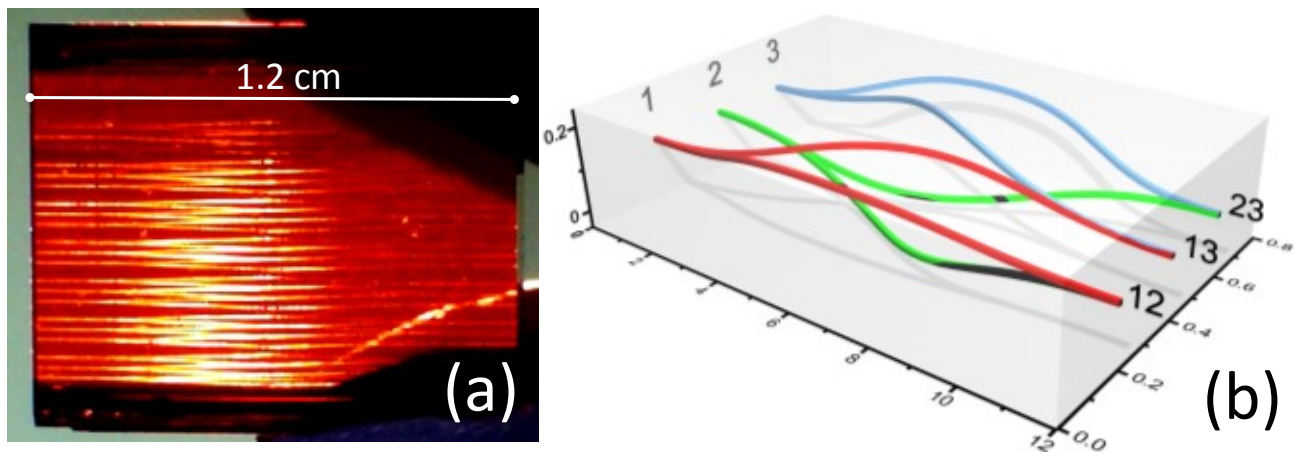

Fig. 2. (Color online) (a) Photograph of our first component in GCIS glass with 10 different 3D 3-beam combiner prototypes for 10.6 μm operation. (b) Sketch of a second 3D 3-beam combiner design with 50 mm and 100 mm bend radii. All units are in mm.

For waveguide DLW, a 1047 nm wavelength, circularly polarized sub-picosecond pulse laser was used, as described elsewhere [13]. Pulse repetition rates were investigated between 0.1 to 1 MHz, with pulse duration in the range 350-460 fs. Writing speeds were studied between 0.25 to 40 mms$^{-1}$, and laser pulse energies from 5 to 50 nJ inside the sample. Focusing aspheric lenses were used with numerical apertures (NAs) of 0.67 to 0.25, with the aim of obtaining different size multiscan waveguide cores, with mean sizes ranging from 2 to 40 μm, and scan spacings from 0.1 to 1 μm, as previously explained elsewhere [13]. Monochromatic guiding characterization of the fabricated waveguides was performed by in- and out-coupling 3.39 μm (HeNe) and 10.6 μm ($CO_2$) laser light by using different ZnSe lenses, as previously described elsewhere [13,9]. Near-field mode imaging of all device outputs was performed using forward looking infrared (FLIR) cameras. Two beam interferometric tests were performed in both monochromatic (10.6 μm) and broadband (3-11 μm) configurations using a Michelson interferometer setup coupled to the $CO_2$ laser and a black body source, as described elsewhere [9].

After extensive waveguide fabrication experiments overall optimum fabrication conditions were identified for each material, which were 0.5 MHz (1.0 MHz), 460 fs (350 fs) and 25 mms$^{-1}$ (30 mms$^{-1}$) writing speeds for GCIS (GLS) samples. This parameter optimization was based on three main factors: (i) maximization of straight waveguide output photon flux at 3.4 μm wavelength, (ii) absence of observable damage under visible light microscope characterization, and (iii) preservation of core rectangular quality and shape. The former feature allowed to perform numerical calculations of near-field modal shape and core index change by supposing ideal step-index rectangular cores and using commercial software COMSOL as described elsewhere [13].

Ten different straight waveguides were fabricated in GCIS glass using different laser powers (fluences) ranging from 19.2 mW (0.16 J/cm$^2$) to 30 mW (0.25 J/cm$^2$) in 1.2 mW steps, and designed to have around 35 μm of horizontal cross-section core size (hereafter waveguides E1 to E10). No guiding at 10.6 μm was observed for channel waveguides fabricated with powers below 20 mW (E1). The horizontal fundamental mode field diameters (MFD) at 1/e$^2$ intensity were measured to range between 58 to 43 μm for this laser power range and core size, at 10.6 μm wavelength. For waveguides written with higher powers than 25 mW a first higher order mode was also observed as shown in Fig. 1(h) for E8 waveguide. From the experimental modes the core index changes could be estimated as previously mentioned. The index change was observed to vary linearly from 0.008 to 0.012 (±10$^{-3}$) for laser fluences ranging from 0.17 to 0.22 J/cm$^2$ (E2 to E6), and to saturate at about $\Delta n$=0.012±10$^{-3}$ for higher fluences (E6 to E10). These values of index change matched both the fundamental mode profiles and the single to bi-modal transition behaviour of the different fabricated waveguides cores.

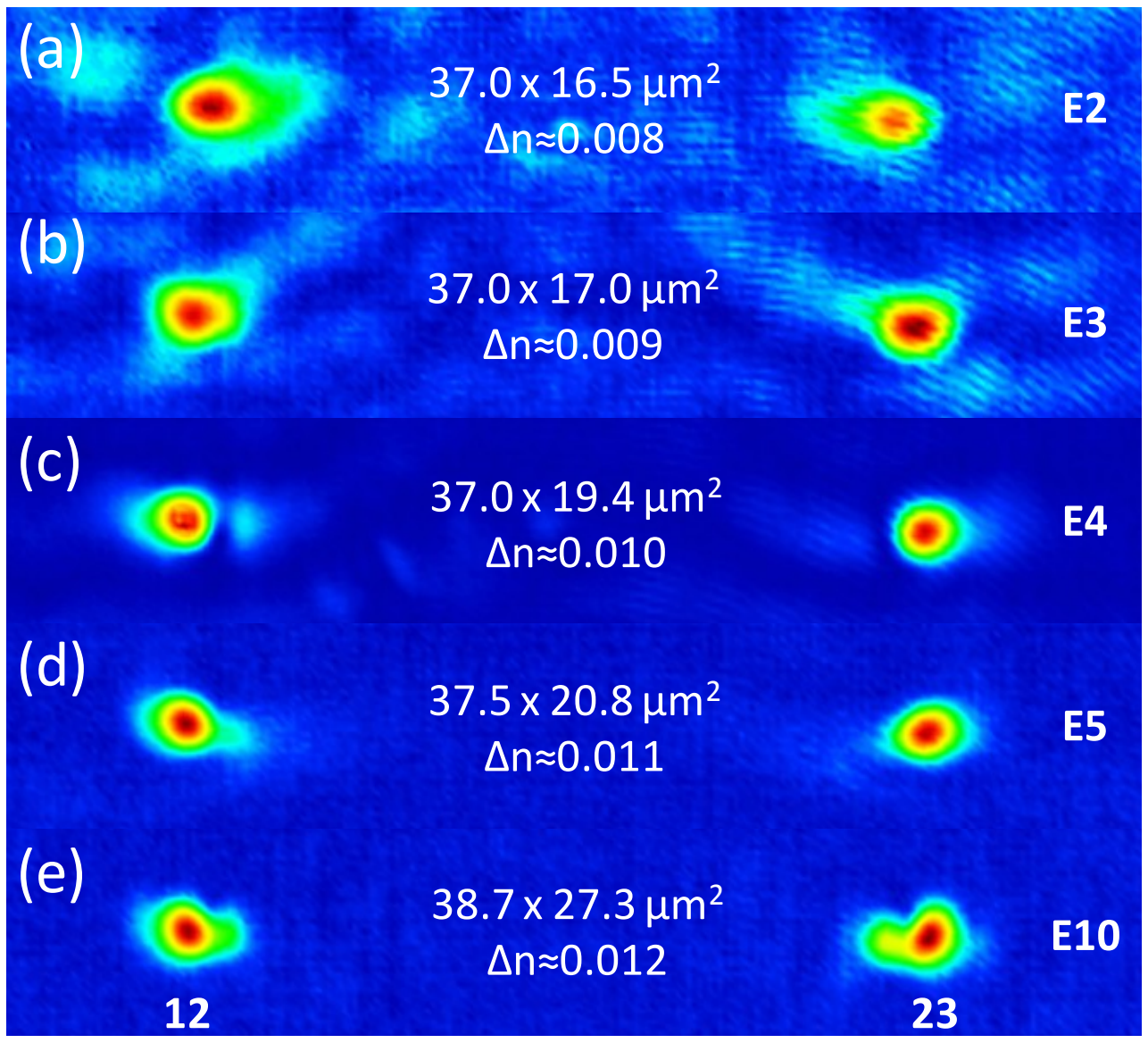

Fig. 3. (Color online) MIR 10.6 μm near-field mode outputs after input 2 injection. The core-size and index-change of the GCIS glass combiners is shown.

Figure 2(a) shows a photograph of the first fabricated prototype 3D three-beam combiner fabricated in GCIS chalcogenide glass in which zero cross-talk waveguide cross-overs are made by vertically separating the channels 200 μm from core centre to centre. The combiner is based on Y-junctions and combines the light from three different inputs (1, 2 and 3), into three different outputs (12, 13, and 23). A sketch of a second prototype 3D beam combiner design is shown in Fig. 2(b). In this case 100 mm waveguide bend radii were implemented for all waveguides, except for 1-12 and 3-23 paths, which had reduced 50 mm radii. All waveguide paths were designed to have the same lengths to within the ±20 nm resolution of the translation stages. Fig. 3 shows the evolution of the GCIS combiners near-field outputs 12 and 23 after input 2 injection, at 10.6 μm wavelength, for five different core parameters (E2 to E5 and E10).

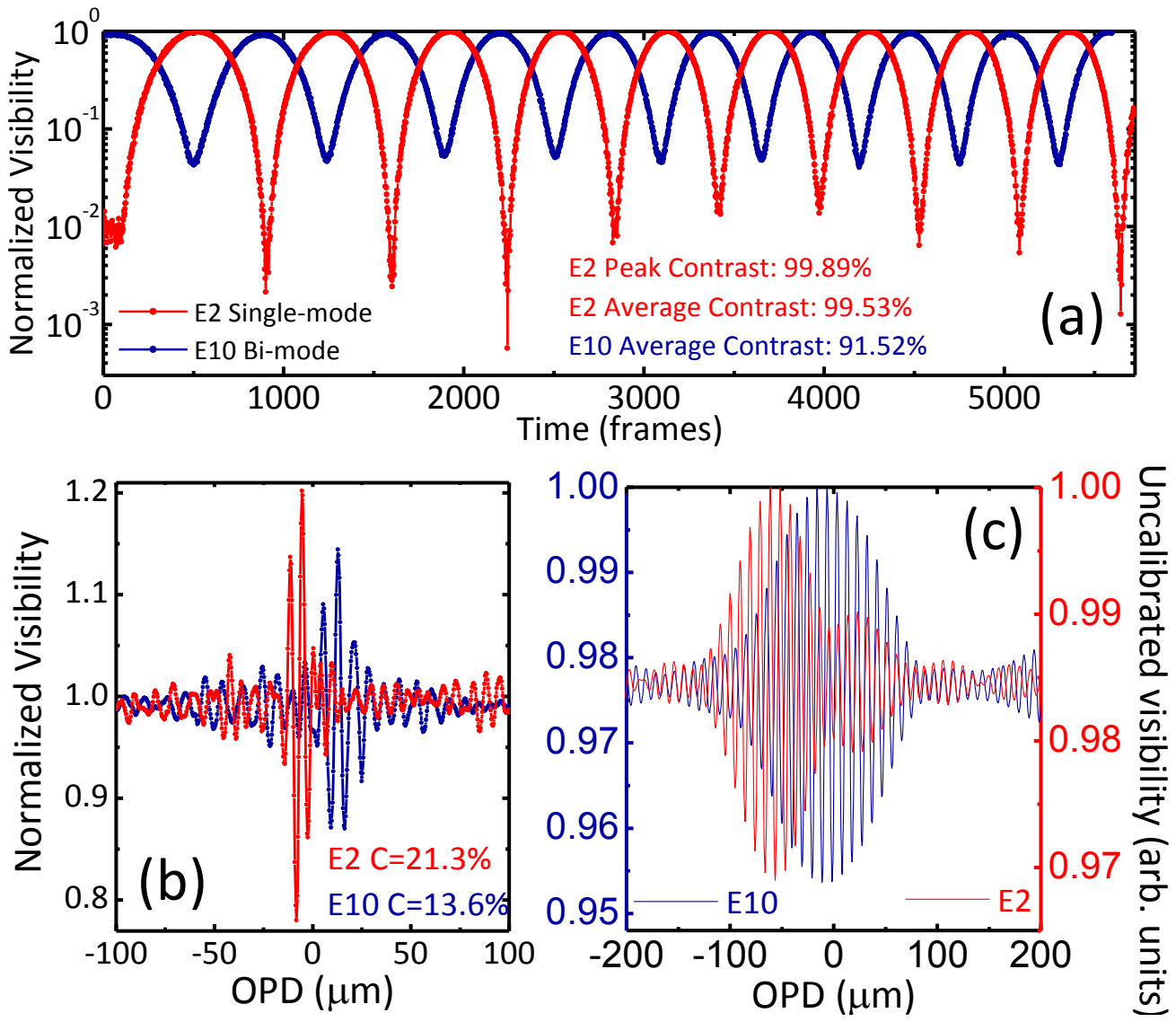

Fig 4. (Color online) (a) Monochromatic visibility at 10.6 μm as a function of time for 100 μm variations of total OPD. (b) Visibility in the region 3-11 μm. (c) Post-processed uncalibrated visibility for the 9-11μm range

Interferometric results performed on the second prototype combiners in GCIS glass are shown in Fig. 4. Fig. 4(a) demonstrates that a monochromatic interference visibility of 99.89% at 10.6 μm wavelength could be achieved with strictly single mode combiners. This level of instrumental contrast confirms the excellent spatial filtering properties of our new 3D devices for stellar interferometers prototyping [17]. Broadband interference was also obtained in the 3-11 μm range to assess the interferometric functioning of the prototype design and technology. Interferometric fringes yielded contrasts around 21% as shown in Fig. 4(b) over the whole spectral band of the test bench (3-11 μm). The decrease in comparison to the monochromatic case is naturally explained by the wide spectral nature of the measurement. Analytical spectral filtering was then performed for the 9-11 μm range (Fig. 4(c)), corresponding to the single-mode regime of the E2 component. This windowing allows visualizing some dispersion effects of the combiners, which are evidenced by the deformation of the fringe package that exhibits phase effects. The origin of these effects can be due to polarization or chromatic behaviours for single-mode regime and also to higher modes influence for the multimode regime. In this later case the dispersion is evidenced by the enlarged fringe package for E10 compared to the expected one for this spectral range for the E2 case. Upcoming measurements will address these issues, especially for various waveguide structures.

In summary, we demonstrate that it is possible to harness the whole mid-infrared transparency range offered by chalcogenide sulphide glasses on a single on-chip photonic instrument, fabricated by means of ultrashort pulse direct laser writing. This finding opens a novel technological avenue to new science in a very wide range of mid-infrared applications. Near future developments will be focused on more specific properties such as accurate contrasts, coupling ratios, effective propagation losses, chromaticity, as well as on the development of more advanced three-dimensional combination schemes.

Authors want to acknowledge J. B. Le Bouquin for fruitful discussion of interferometric data reduction. AR acknowledges financial support from the Spanish Ministerio de Educación under the Programa de Movilidad de Recursos Humanos del Plan Nacional de I+D+I 2008/2011 for abroad postdoctoral researchers. RRT acknowledges support through an STFC Advanced Fellowship (ST/H005595/1). This work was also funded by the UK EPSRC grants EP/F067690/1, EP/G030227/1 and EP/D048672/1, and by the ongoing Indo-UK collaboration under the UK-India Education and Research Initiative (UKIERI).